\documentclass[10pt,conference]{IEEEtran}

\usepackage{graphics} 
\usepackage[pdftex]{graphicx}
\usepackage{times} 
\usepackage [cmex10]{amsmath} 
\usepackage{amssymb}  
 \IEEEoverridecommandlockouts
\title{
Joint Source-Channel Coding  over a  Fading Multiple Access Channel
with Partial Channel State Information
\thanks{This work was partially supported by the DRDO-IISc program on Advanced Research in Mathematical Engineering.}}
\author{ R Rajesh and Vinod Sharma\\
 Dept. of Electrical Communication Engineering,\\
  Indian Institute of Science, Bangalore, India\\Email:
rajesh@pal.ece.iisc.ernet.in,~vinod@ece.iisc.ernet.in}
\begin{document}
\maketitle \thispagestyle{empty} \pagestyle{empty}
\begin{abstract}
In this paper we address the problem of transmission of correlated sources over a fast fading multiple access channel (MAC) with partial channel state information available at both the encoders and the decoder. We provide sufficient conditions for transmission with given distortions. Next these conditions are specialized to a Gaussian MAC (GMAC). We provide the optimal power allocation strategy and compare the strategy with various levels of channel state information.\\
Keywords: Fading MAC, Power allocation, Partial channel state information, Correlated sources.

\end{abstract}
\section{Introduction and survey}
\label{intro}
Sensor networks are used in a wide variety of applications due to their ability to operate in environments where human penetration is not
possible. These networks are characterized by inexpensive sensing nodes with limited battery power and storage and hence limited computing and communication capabilities~\cite{akylidiz02survey}. The  sensor nodes may  often be deployed for monitoring a random field. Due to the spatial proximity of the nodes, sensor observations are correlated. One often needs to transmit these observations to a fusion center through wireless links which experience multipath fading. The encoders and decoder often have only a partial information about the channel state as it has to be learnt. This happens because the estimate of the channel state will often not be accurate. But some estimate of the channel state may be available from 'Hello' messages passed.  A fundamental building block for such a network is a fading  Multiple Access Channel (MAC) with partial channel state information at the encoders and the decoder.  We study such a system in this paper.

In the following we survey the related literature. Cover, El Gamal and Salehi~\cite{Cover80multiple} provided sufficient conditions for transmitting losslessly discrete correlated observations over a discrete MAC. They also show that unlike for independent sources, the source-channel separation does not hold. These techniques were extended to more general models with discrete sources and channels and lossless transmission in \cite{Ahlswede83source}. Reference~\cite{Rajesh07gaussian} extends the result in~\cite{Cover80multiple} and  obtains sufficient conditions for lossy transmission of correlated sources over a MAC with side information. 

The multi-access fading channels with independent inputs were  considered in the excellent survey  \cite{proakis}. They  show that unlike in the single user case, in the multi-access realm the optimal power control yields a substantial gain in capacity. The optimal power allocation strategy for the symmetric case is to allow the  user with the best channel to transmit at a time (Random TDMA) (\cite{knopp1}). The instantaneous power allocated to a user is by the well known `water-filling' algorithm in  time.

Multiple-access techniques for fading cellular uplink model with adjacent cell interference are discussed in \cite{shamai}.

Das and Narayan  \cite{das} obtained the capacity for a time varying MAC with varying degrees of channel state information at the encoders and the decoder as a union of  regions characterised by limits of mutual information. A single letter specialization of this result to a fading memoryless channel with stationary ergodic state process, perfect CSIR and partial CSIT is given in \cite{now}. The case with perfect CSIT  and no CSIR is studied in \cite{isit05}. 

The capacity of a fading Gaussian channel with channel state
information (CSI) at the transmitter and receiver and at the
receiver alone are provided in \cite{varia}. It was shown that
optimal power adaptation when CSI is available both at the
transmitter and the receiver is `water filling' in time. The
capacity region of the Gaussian MAC (GMAC) with independent inputs
is available in \cite{Cover04elements}. The distributed Gaussian
source coding problem is discussed
in~\cite{Oohama97Gaussian,Wagner05rate}. Exact rate region {for two
users} is provided in \cite{Wagner05rate}. Joint source-channel coding schemes for transmission of correlated sources over a  Gaussian MAC are  also discussed in \cite{Gastpar05power} and \cite{Lapidoth01sending}.
Hanly and Tse
(\cite{tse}) have generalized the results on a GMAC with independent
inputs to a  flat fading GMAC with  perfect CSI at both the
transmitters and the receiver. Gaussian MAC with independent sources, partial CSIT and perfect CSIR is studied in  \cite{isit02} for lossless transmission.

An explicit characterization of the ergodic capacity region and a
simple encoding-decoding scheme for a fading GMAC with common data
is given in \cite{nan} for lossless transmission. Optimum power
allocation schemes are also provided.

Joint source channel coding for correlated data on a fading MAC with perfect CSI at the transmitters (CSIT) and  perfect CSI at the receiver (CSIR) is studied in \cite{allerton08}. Various power allocation schemes are studied and it is shown that the Random TDMA scheme is suboptimal for transmission of correlated sources. Optimal power allocation strategies to minimize the sum distortion for transmission of discrete and Gaussian sources over a GMAC are given.

This paper makes the following contributions. Sufficient conditions for lossless and lossy transmission of correlated sources over a fading MAC with partial state information available at the encoders and the decoder are obtained. The source alphabet and/or the channel alphabet can be discrete or continuous. The conditions are specialized for GMAC and an optimal power allocation policy is derived. The performace of the power allocation policy for various levels of channel state information is studied. The  power allocation policies are used in combination with joint sources-channel codes to transmit correlated sources over fading channels. Various previous results are shown as special cases.

 The rest of the paper is organized as follows. Sufficient conditions for transmission of correlated sources over a fading MAC with partial CSIT and partial CSIR are provided in  Section II. The special cases are provided in Section III. These conditions are specialized to a fading GMAC in Section IV.  Section V obtains the optimal power allocation policy.  Section VI concludes the paper. Proof of the main result is given in the Appendix.

\section{Transmission of correlated sources over a fading MAC with partial CSIT and partial CSIR}\label{sec1.3}
In this section we consider the transmission of memoryless dependent
sources, through a memoryless fading multiple access channel. The
sources and/or the channel input/output alphabets can be discrete or
continuous. The transmitters  and the receiver
have partial knowledge of the fade state of the channel at that
time.

We consider two sources $(U_1,U_2)$ with a known joint distribution
$F(u_1, u_2)$. The random vector sequence $\{(U_{1n}, U_{2n}),~
n\geq1\}$ formed from the source outputs with distribution $F$ is
independent identically distributed ({\it{iid}}) in time. The
sources transmit their code words $X_i$'s to a single decoder
through a memoryless, flat, fast fading multiple access channel. Let
$(H_{1n},H_{2n})$ be the fade state at time $n,~ n\geq1$. We assume
$\{(H_{1n},H_{2n}),~n\geq 1\}$ to be an $iid$ sequence, although
$(H_{1n},H_{2n})$ can be dependent and can be discrete or continuous
valued. Similarly $(\hat{H}_{1n},\hat{H}_{2n})$,
$(\tilde{H}_{1n},\tilde{H}_{2n})$ denote the channel state information
available at the transmitters and the receiver respectively.
$(H_{1n},H_{2n},\hat{H}_{1n},\hat{H}_{2n},\tilde{H}_{1n},\tilde{H}_{2n})$
are $iid$ in time. The channel output $Y$ has distribution
$p(y|x_1,x_2,h_1,h_2)$ if $x_1$ and $x_2$ are transmitted at that
time  and the channel is in the fade state $(h_1,h_2)$.  The decoder
receives $Y$ and  estimates the sensor observations  $U_i$   as $
\hat{U_i},~i=1, 2$.

It is of interest to find encoders and a decoder such that
$\{U_{1n},U_{2n},n\geq1\}$ can be transmitted over the given  fading
MAC with $E[d_i(U_i,\hat{U_i})]\leq D_i,~i=1,2$  where $d_i$  are
non-negative distortion measures and $D_i$ are the given distortion
constraints. Encoder $i$ knows $U_{in}$ and $(\hat{H}_{1n},\hat{H}_{2n})$ at time $n$. This situation corresponds to the scenario where the channel state is broadcasted to the encoder via a third party by means of wayside channels (see \cite{cem}). We will assume that $d_i$ are such that $d_i(u,u')=0$
if and only if $u=u'$. If the distortion measures are unbounded we
also assume that there exist $u_i^*,~i =1, 2$ such that
$E[d_i(U_i,u_i^*)] < \infty,~i = {1, 2}$. This condition is satisfied in the important special case of mean square error if we take $u_i^*$ to be $E[U_i]$ and $E[U_i^2] < \infty$.

Due to correlated sources, source channel separation does not hold in this case.

We will denote $U_{ij},j=1,2,...,n$  by $U_i^n, i=1,2$. Also $x$ will denote a realization of a random variable $X$.

$Definition$ : The sources $(U_1^n ,U_2^n)$ can be transmitted over
the fading  multiple access channel with distortions
${\bf{D}}{\buildrel\Delta \over=}(D_1,D_2)$ if for any $\epsilon  >
0 $ there is an $n_0$ such that for all $n > n_0$  there exist
encoders $f_{E,i}^n: \mathcal{U}_i^n \times \hat{\mathcal{H}}_1^n
\times \hat{\mathcal{H}}_2^n \rightarrow \mathcal{X}_i^n ,~i =1,2$
and a decoder ${{f}_{D}^n}: \mathcal{Y}^n \times \mathcal{\tilde{H}}_1^n
\times \mathcal{\tilde{H}}_1^n \rightarrow (\mathcal{\hat{U}}_1^n ,
\mathcal{\hat{U}}_2^n) $   such that $
\frac{1}{n}E\left[\sum_{j=1}^nd (U_{ij},\hat{U}_{ij})\right]\leq
D_i+ \epsilon,~i=1,2$ where $(\hat{U}_1^n,\hat{U}_2^n )=
f_D(Y^n,\tilde{H}_1^n,\tilde{H}_2^n)$ and  $\mathcal{U}_i,~\mathcal{H}_i,~\hat{\mathcal{H}}_i,~\tilde{\mathcal{H}}_i,
~\mathcal{X}_i,~\mathcal{Y},~\hat{\mathcal{U}_i} $  are the sets in
which $ U_i,~H_i,~\hat{H}_i,~\tilde{H}_i,~X_i,~Y$ and $\hat{U}_i$
take values.

Since the MAC is memoryless,
$p(y^n|x_1^n,x_2^n,h_1^n,h_2^n)=\prod_{j=1}^n
p(y_j|x_{1j},x_{2j},h_{1j},h_{2j})$. In the following
$X\leftrightarrow Y \leftrightarrow Z$   will indicate that $(X, Y,
Z)$ forms a Markov chain.

Now we state the main Theorem. \\
    \newtheorem{theorem}{Theorem} 
    \begin{theorem}
A source  can be transmitted over a fading  multiple access channel
with distortions $(D_1, D_2)$ if there exist random variables $(W_1,
W_2,  X_1,  X_2)$ such  that
\begin{multline}
 1.~ p(h_1,h_2,\hat{h}_1,\hat{h}_2,\tilde{h}_1,\tilde{h}_2,u_1,u_2,w_1,w_2,x_1,x_2,y) =\\\nonumber p(h_1,h_2,\hat{h}_1,\hat{h}_2,\tilde{h}_1,\tilde{h}_2)p(u_1,u_2) p(w_1|u_1)  p(w_2|u_2)\nonumber\\
      p(x_1|w_1,\hat{h}_1,\hat{h}_2)p(x_2|w_2,\hat{h}_1,\hat{h}_2)p(y|x_1,x_2,h_1,h_2). \nonumber
\end{multline}

$2.$ There exists a function $f_D: \mathcal{W}_1 \times
\mathcal{W}_2 \times \mathcal{\tilde{H}}_1 \times \mathcal{\tilde{H}}_2\rightarrow
(\hat{\mathcal{U}}_1 \times  \hat{\mathcal{U}}_2) $  such that $
E[d(U_{i},\hat{U}_{i})]\leq D_i,~i=1,2$ and the constraints
 \begin{eqnarray}
\label{constraints}
 I (U_1; W_1 | W_2)  &<&  I (X_1; Y | X_2, W_2,\tilde{H}_1,\tilde{H}_2),\nonumber\\
 I (U_2; W_2 | W_1) &<& I (X_2; Y | X_1, W_1,\tilde{H}_1,\tilde{H}_2),\\
  I (U_1,U_2 ; W_1, W_2) &<& I (X_1, X_2; Y|\tilde{H}_1,\tilde{H}_2),\nonumber
\end{eqnarray}
are satisfied where $\mathcal{W}_i$ is the set in which $W_i$ take
values.~~~~~~~~~~~~~~~~~~~~~~~~~~~~~~~~~~~~~~~~~~~~~~~~~~~~~~~~~~~~~~~~~~~~~~~~~\QED

If the channel input alphabets are continuous valued then the $X_is$
should also satisfy given power constraints $E[X_i^2]\leq
\overline{P}_i,~i=1,2$. \label{theorem1}
\end{theorem}

\begin{figure*}[t]
\centering
{\scriptsize \parskip=9pt{
\begin{eqnarray}
\label{constraintsg}
I (U_1; W_1 | W_2)  &<&  0.5E\left[\log\left( 1+ \frac{|H_1|^2P_1(\hat{H}_1,\hat{H}_2)(1- {\tilde{\rho}}^{2})} {{\sigma_N}^{2}}\right)\right]\nonumber,\\
I (U_2; W_2 | W_1)  &<&  0.5E\left[\log\left( 1+ \frac{|H_2|^2P_2(\hat{H}_1,\hat{H}_2)(1- {\tilde{\rho}}^{2})} {{\sigma_N}^{2}}\right)\right],\\
I (U_1, U_2 ; W_1, W_2)  &<&   0.5E\left[\log\left( 1+
\frac{|H_1|^2P_1(\hat{H}_1,\hat{H}_2) +
|H_2|^2P_2(\hat{H}_1,\hat{H}_2) + {2}|H_1||H_2|
{\tilde{\rho}}{\sqrt{P_1(\hat{H}_1,\hat{H}_2)P_2(\hat{H}_1,\hat{H}_2)}}}
{{\sigma_N}^{2}} \right)\right]\nonumber.
\end{eqnarray}}}
\hrulefill
\end{figure*}


The proof of the theorem is given in the Appendix.
The proof extends directly to the multi-user case (with the number of users $>2$). Let $\mathcal{S}={1,2,...,M}$ be the set of sources with joint distribution $p(u_1,u_2,...,u_M)$. $H_{\mathcal{S}}$ denotes the set $\{H_1,H_2,...,H_M\}$. Similarly we define $\hat{H}_{\mathcal{S}}$ and $\tilde{H}_{\mathcal{S}}$ for the CSIT and CSIR respectively.

\begin{theorem}
Sources $(U_i^n ,i\in \mathcal{S})$  can be communicated in a distributed fashion over the memoryless fading multiple access channel $p(y|x_i ,i\in \mathcal{S})$  with distortions $(D_i ,i\in \mathcal{S})$  if there exist auxiliary random variables $(W_i,X_i ,i\in \mathcal{S})$ satisfying
\begin{eqnarray*}
1.~ p(u_i,w_i,x_i,y,h_i,\tilde{h}_i,\hat{h}_i, i \in \mathcal{S}) =
     p(h_i,\tilde{h}_i,\hat{h}_i,i \in \mathcal{S})\\
  p(u_i,i \in \mathcal{S})p(y|x_i,h_i,i \in \mathcal{S})\prod_{j \in \mathcal{S}}p(w_j|u_j)p(x_j|w_j,\hat{h}_{\mathcal{S}}).
\end{eqnarray*}
$2.$ There exists a function $f_D: \prod_{j \in \mathcal{S}}\mathcal{W}_j \times  \mathcal{\tilde{H}} \rightarrow (\hat{\mathcal{U}}_i,i \in \mathcal{S}) $  such that $ E[d(U_{i},\hat{U}_{i})]\leq D_i,~i \in \mathcal{S}$ and the constraints
\begin{equation*}
I (U_A; W_A | W_{A^c}) < I (X_A; Y | X_{A^c}, W_{A^c}, \tilde{H}_{\mathcal{S}})
\label{eqn1.3}
\end{equation*}
are satisfied  for all $ A  \subset \mathcal{S}$ (in case of continuous channel alphabets we also need the power constraints $E[X_i^2] \leq \overline{P}_i,~i=1,...,M)$.~~~~~~~~~~~~~~~~~~~~~~~~~~~~~~~~~~~~~~~~~~~~~~~~~~~~~~~\QED
\end{theorem} 

In Theorem~\ref{theorem1} it is possible to include other distortion
constraints. For example, in addition to the bounds on
$E[d(U_i,\hat{U}_i)]$ one may want a bound on the joint distortion
$E[d((U_1,U_2),(\hat{U}_1,\hat{U}_2))]$. Then the only modification
needed in the statement of the above theorem is to include this also
as a condition in defining $f_D$.

\section{Special Cases}
In the following we show that our results contain several previous studies as special cases.
\subsection{Lossy transmission of correlated sources with perfect CSIT, perfect CSIR}
Take $(\hat{H}_1,\hat{H}_2)=(\tilde{H}_1,\tilde{H}_2)= (H_1,H_2)$,
then we recover the conditions given in \cite{allerton08}.
\subsection{Lossless transmission of independent sources with partial CSIT, perfect CSIR}
Take $(U_1,U_2)$ as discrete valued and $(\tilde{H}_1,\tilde{H}_2)=(H_1,H_2)$. Also, ${\bf{H}}=(H_1,H_2)$, $U_1 \bot U_2$ (i.e., $U_1$ is independent of $U_2$). We also take $(W_1,W_2)=(U_1,U_2)$. Then
we need to code at rate $R_1$ and $R_2$ satisfying

 \begin{eqnarray*}
\label{constraintsn}
H(U_1)< R_1  &<&  I (X_1; Y | X_2 ,\bf{H}),\nonumber\\
 H(U_2) < R_2  &<& I (X_2; Y | X_1,\bf{H} ),\\
  H(U_1U_2)< R_1+R_2 &<& I (X_1, X_2; Y|\bf{H}).\nonumber
\end{eqnarray*}
In the above equations $X_1$ and $X_2$ are generated using $(\hat{H}_1,\hat{H}_2)$ and are independent of $(U_1,U_2)$.

These  are the conditions given in \cite{now} and specializations of the general conditions in \cite{das}.

\subsection{Lossless transmission of independent sources with Perfect CSIT, no CSIR}
Take $(\hat{H}_1,\hat{H}_2,\tilde{H}_1,\tilde{H}_2)=(H_1,H_2,1,1)$
and $(W_1,W_2)=(U_1,U_2)$. We consider no channel state information
at the decoder and perfect channel state information at the
encoders. Then one needs to code at rates $R_1$ and $R_2$ satisfying

 \begin{eqnarray*}
\label{constraintsmm}
H(U_1)< R_1  &<&  I (X_1; Y | X_2 ),\nonumber\\
 H(U_2) < R_2  &<& I (X_2; Y | X_1),\\
  H(U_1U_2)< R_1+R_2 &<& I (X_1, X_2; Y).\nonumber
\end{eqnarray*}

These are the conditions in \cite{isit05}.


\section{Fading Gaussian MAC with partial CSIT, perfect CSIR}
In a fading Gaussian MAC the channel output $Y_n$ at time $n$ is
given by $Y_n = H_{1n}X_{1n} + H_{2n}X_{2n} + N_n$  where $X_{1n}$
and $X_{2n}$  are the channel inputs at time $n$ and $\{N_n\}$  is
iid with a Gaussian distribution and is independent of $X_{1n}$  and
$X_{2n}$. Also, $E[N_n] =0 $ and $var(N_n)= \sigma_N^{2}$. $H_{1n}$
and $H_{2n}$ are the fade states of the channel at time $n$.  The
power constraints on the channel inputs are $E[X_i^2] \leq
\overline{P}_i,~i=1,2$. The distortion measure will be Mean Square
Error (MSE). Let $\tilde{\rho}$ be the correlation between the
channel inputs $X_1,X_2$.

For this GMAC, following the experience in  \cite{bitg} we relax
the first two inequalities in \eqref{constraints} to make them more
explicit. These are then used to obtain efficient signaling schemes
to satisfy \eqref{constraints}. For this the right hand side (RHS) of the first two
inequalities in \eqref{constraints} are replaced by upper bounds
$I(X_1;Y|X_2,H_1,H_2)$ and $I(X_2;Y|X_1,H_1,H_2)$  respectively.  It
is shown in  \cite{bitg} that these upper bounds are quite tight
whenever these two inequalities are active (generally it is the
third inequality which is tight). Also, it is shown in
\cite{bitg} that for a given $(h_1,h_2)$, these upper bounds and
the RHS of the third inequality in \eqref{constraints} are
maximized by choosing $(X_1,X_2)$ to be zero mean, jointly Gaussian
r.v.s with $E[X_i^2]=P_{i}(\hat{h}_1,\hat{h}_2)$ where $P_{i}(\hat{h}_1,\hat{h}_2)$ are appropriately chosen. If  such
$(X_1,X_2)$ have correlation $\tilde{\rho}$ these three bounds
provide \eqref{constraintsg}.

The expectation is over the joint fade state
$(h_1,h_2,\hat{h}_1,\hat{h}_2)$. We also need to choose the power
control policies $P_{i}(\hat{h}_1,\hat{h}_2)$  such that the average
power constraints
\begin{equation}
E[P_i(\hat{H}_1,\hat{H}_2)]\leq \overline{P}_i,~i=1,2,
\end{equation}
are satisfied. This motivates us to consider Gaussian coding
schemes.

An advantage of \eqref{constraintsg}  is that we will be able to
obtain explicit source-channel coding schemes to satisfy
\eqref{constraintsg}. These may be difficult to identify from
\eqref{constraints} itself. Once we have obtained these coding
schemes we can verify the sufficient conditions \eqref{constraints}
themselves. If satisfied these will ensure that the coding schemes
can ensure transmission with given distortions. If not, one can
change $\tilde{\rho}$ to finally satisfy \eqref{constraints}. Thus
in the rest of the paper we consider some power allocation policies
along with Gaussian signaling schemes which can be used to satisfy
the conditions \eqref{constraints}.

\subsection {Special Case}
Specializing the theorem for independent sources and lossless
transmission recovers the result in \cite{isit02}.


\section{Optimal power allocation for the GMAC}

We  consider a power  allocation policy such that the RHS
in the third inequality of \eqref{constraintsg}   is maximized and the
other conditions are satisfied. This is done because often the third
inequality is the constraining condition. In \cite{allerton08} this policy was compared with sevaral other power allocation policies and was found to perform better. This optimal power
allocation policy is obtained numerically. It depends upon
$\tilde{\rho}$ which in turn depends on the source correlation
$\rho$.

To find a $\tilde{\rho}$ such that all the three inequalities are
satisfied by the optimal policy, we can use the following procedure.
We consider an iterative algorithm in which the channel input correlation
$\tilde{\rho}$ is chosen in such a way that the third inequality
is satisfied. Then we check for the other two inequalities. If they
fail then the $\tilde{\rho}$ is decreased so that all the three
conditions are satisfied, if possible. These computation need to be done only once for a given source pair distribution. Also, the computational requirements are modest.

One of the performance measures is to maximize the RHS in the
third inequality of \eqref{constraintsg}. The optimal power
allocation policy discussed in the beginning of this section
maximizes this. We take  $\overline{P_1}=\overline{P_2}=1$. The
channel correlation ($\tilde{\rho}$) achievable depends on the
source correlations and the scheme used. The fade states $h_1,h_2$
take values in $(1,0.5)$ with equal probability and are independent
of each other. The channel noise is zero mean with unit variance.
The channel state information at the transmitter (CSIT) is a noise
corrupted version of the fade state. Thus the CSIT can be modeled as
the output of a binary symmetric channel (BSC) with fade states as
the input and with a suitable cross over probabilty $p$. The results provided are exact computations of RHS of \eqref{constraintsg}. 

\begin{figure}[h]
\centering
\includegraphics [height=2.5in, width=3in ] {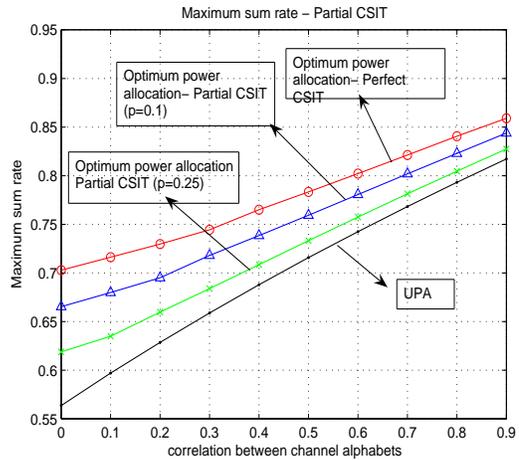}
\caption{Maximum sum rate with partial CSIT,
$\overline{P_1}=\overline{P_2}=1$} \label{fig_fade}
\end{figure}

We compare the performance of the optimum power allocation that
maximizes the RHS in the third inequality of \eqref{constraintsg}
for various values of CSIT. Figure \ref{fig_fade} compares the cases
with perfect CSIT ($p=0$), partial CSIT ($p=0.1,0.25$) and no CSIT
($p=0.5$). For the no CSIT case uniform power allocation (UPA) is the optimal power allocation
policy as the objective function is monotonic in both $P_1$ and
$P_2$.

We observe that uniform power allocation (UPA) performs well at high $\rho$ and is comparable to the
optimum power allocation when $p$ is close to 0.5.

\begin{figure}[h]
\centering
\includegraphics [height=2.5in, width=3in ] {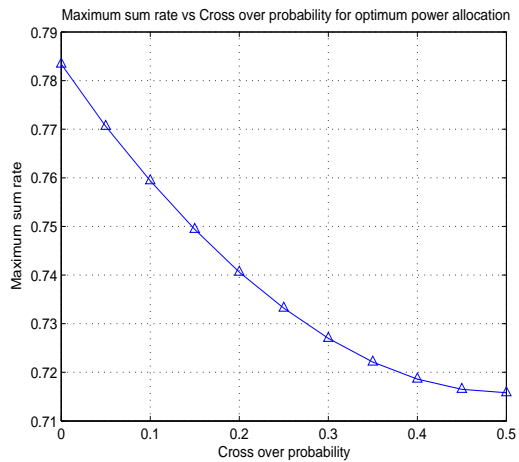}
\caption{Maximum sum rate,
$\overline{P_1}=\overline{P_2}=1,~\tilde{\rho}=0.5$}
\label{fig_fade1}
\end{figure}

Figure \ref{fig_fade1} shows the variation in the maximum sum rate
for various $p$.

\subsection{Discrete sources}\label{sec6a}
Next we obtain the optimum power allocation for transmitting
correlated discrete sources over a GMAC with various levels of
partial CSIT. Consider the lossless transmission of discrete sources
$(U_1,U_2)$ over a fading GMAC. Such a system is most commonly
encountered in practice. We consider the example where the sources $(U_1,U_2)$ have joint
distribution given by $P(U_1=0; U_2=0) = P(U_1=1; U_2=1)= P(U_1=0;
U_2=1)=1/3;P(U_1=1; U_2=0)=0$. The fade states $h_1,h_2$ take values
in $(1,0.5)$ with equal probability and are independent of each
other.  The power constraints on the channel inputs $(X_1,X_2)$ are
$(5,5)$. The channel noise is zero mean with unit variance.

For lossless transmission the left hand side (LHS) in \eqref{constraintsg} become
$H(U_1|U_2),~ H(U_2|U_1)$  and  $H(U_1,U_2)$ respectively and they
evaluate to 0.667, 0.667 and 1.585. Let the sources be mapped to
channel codewords with correlation $\tilde{\rho}=0.3$.  Such
correlation preserving mappings are discussed in  \cite{bitg}. If we using UPA the RHS in the third inequality evaluates  to 1.5030 and the sources cannot be transmitted over the channel losslessly.
 Using the optimum power allocation scheme for perfect CSIT and CSIR, RHS
evaluates to  1.6071. The RHS in the first two inequalities
evaluate to  0.8755 with the optimal policy. This ensures that the
sources can be transmitted  losslessly. 

  Now we consider partial CSIT and perfect CSIR for the
 above example. The partial CSIT is modeled as the output of a BSC with
 suitable crossover probability $p$ as shown above. For $p=0.1$ the RHS in the third inequality is 1.585 and the first two inequalities are 0.8650. Hence with this partial CSIT, we are able to transmit the given sources over the GMAC losslessly.

\subsection{Gaussian Sources}
Consider the transmission of correlated Gaussian sources over a
GMAC. The sources are assumed to have zero mean, unit variance and
correlation $\rho$. The power constraints on the channel inputs are
$(\overline{P_1},\overline{P_2})$ and channel noise variance is
unity. The performance measure is to find the minimum distortion at
the decoder for  given sources, power constraints and channel noise
variance. The distortion criterion is mean square error (MSE).

In  \cite{Rajesh07allerton} and \cite{allerton08} we have discussed
three joint source channel coding schemes for transmission of
Gaussian sources over a GMAC. It is found that the Lapidoth-Tinguely
(LT) scheme (developed in \cite{Lapidoth01sending}) is a good coding
scheme that performs well over all SNR regions. In this scheme the
sources are vector quantized and mapped to correlated Gaussian
codewords. The decoding is by jointly typical decoding followed by
estimation of the sources.

We use the optimum power allocation scheme obtained above with LT scheme to obtain the distortions for 
different levels of CSIT. We assume perfect CSIR in all
cases. We will take $(\overline{P_1},\overline{P_2}) =(5,5)$ and
$\sigma_N^2=1$. The fading processes are as in Section \ref{sec6a}.

\begin{figure}[h]
\centering
\includegraphics [height=2.5 in, width=3in ]{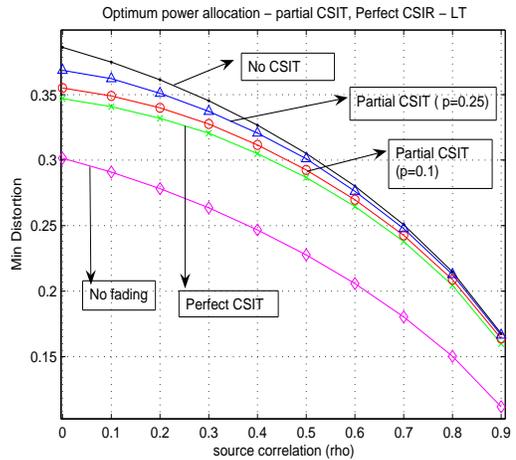}
\caption{Minimum distortion for LT,
$\overline{P_1}=\overline{P_2}=5$} \label{figlt}
\end{figure}

The minimum distortions achieved for $p=0,0.1,0.25$ and  $0.5$ are
plotted in Figure \ref{figlt}. The distortions for the Gaussian sources are also computed from \eqref{constraintsg}  by choosing $W_i$ as vector quantized version of $U_i$ with rate $R_i$ and finding $var[U_i|W_1,W_2],~ i=1,~2$. $p=0$ corresponds to the perfect CSIT
case and $p=0.5$ corresponds to UPA. For comparison purposes the
minimum distortions achieved in a channel without fading are also provided. It is
seen from this figure that the partial CSIT affects the distortions
more at lower $\rho$.  The fade state (1, 0.5) is not that strong a channel value. If the  channel state can get worse, our optimal algorithm will perform  better compared to UPA as the power allocation guarantees allocation of power to good states only.

\section{Conclusions and Discussion}
In this paper, sufficient conditions for transmission of correlated sources over
a fading MAC with partial CSIT and CSIR are provided. These conditions are specialized to
 a GMAC and used to obtain a good power allocation policy. This policy is then used with efficient joint source-channel coding schemes for transmission of discrete and Gaussian sources. Consequently, we have identified very efficient (even though not provably optimum) signalling schemes for a fading GMAC with partial CSIT and CSIR.
 
 The results in this paper are information-theoretic  and hence address the fundamental limitations of the communication system. In this paper, the codes used are of infinite length and the encoders and decoder used provide the best conditions that are feasible.  In practice, one needs to develop practical coding schemes to design  a system that can perform close to the optimal conditions provided here. Also, our results, as often happens in information theory, can provide significant insights in designing practical codes and the power control algorithms. 

\begin{appendix}
\section{Proof of Theorem 1}
First we consider discrete sources and channel inputs. Comments to
include the continuous sources/channel inputs are provided at the
end of the proof.

The scheme involves distributed quantization $(W_1^n,W_2^n)$ of the
sources followed by a correlation preserving mapping to the channel
codewords depending on the channel state. The decoding approach
involves first decoding the quantized version $(W_1^n,W_2^n)$ of
$(U_1^n,U_2^n)$ using $(Y^n,\tilde{H}_1^n,\tilde{H}_2^n)$ and then
obtaining estimate $(\hat{U}_1,\hat{U}_2)$ as a function of
$(W_1^n,W_2^n)$.

We show the achievability of all points in the rate region (1).

$Proof$: Fix $p(w_1|u_1)$, $p(w_2|u_2)$,
$p(x_1|w_1,\hat{h}_1,\hat{h}_2)$, $p(x_2|w_2,\hat{h}_1,\hat{h}_2)$
as well as $f_D$ satisfying the distortion constraints.

$Codebook~ generation$: Let $R_i^{'}=I(U_i;W_i)+\delta,~ i = {1,2}$
for some $\delta >0$. Generate $2^{nR_i^{'}}$ codewords of length
$n$, sampled $iid$ from the marginal distribution $p(w_i),~i ={1,2}$.
For each $w_i^n$ and $(\hat{h}_1^n,\hat{h}_2^n)$ independently
generate sequence $X_i^n$ according to $\prod_{j=1}^n
p(x_{ij}|w_{ij},\hat{h}_{1j},\hat{h}_{2j}),~ i = {1,2}$. Call these
sequences $x_i^n(w_i^n,\hat{h}_1^n,\hat{h}_2^n), i \in {1,2}$.
Reveal the codebooks to the encoders and the decoder.

$Encoding$: For $i \in \{1,2\}$, given the source sequence $U_i^n$
and  $\hat{h}_1^n,\hat{h}_2^n$, the $i^{th}$ encoder looks for a
codeword $W_i^n$ such that $(U_i^n,W_i^n) \in
T_{\epsilon}^n(U_i,W_i)$ and then transmits
$X_i^n(W_i^n,\hat{h}_1^n,\hat{h}_2^n)$ where $T_{\epsilon}^n(.)$ is
the set of weakly $\epsilon$-typical sequences
(\cite{Cover04elements}) of length $n$.

$Decoding$: Upon receiving $Y^n$, for a given
$(\tilde{h}_1^n,\tilde{h}_2^n)$ the decoder finds the unique
$(W_1^n,W_2^n)$ pair such that
$(W_1^n,W_2^n,x_1^n(W_1^n,\overline{h}_1^n,\overline{h}_2^n),x_2^n(W_2^n,\overline{h}_1^n,\overline{h}_2^n),Y^n)\in
T_{\epsilon}^n$, where $\overline{h}_1^n,\overline{h}_2^n$ takes values in $\hat{h}_1^n,\hat{h}_2^n$. If it fails to find such a unique pair, the decoder declares $({u_1^*}^n,{u_2^*}^n)$.

In the following we show that the probability of error for the above
encoding, decoding scheme tends to zero as $n \rightarrow \infty$.
By Markov Lemma (\cite{Cover04elements}),
$P\{(U_1^n,U_2^n,W_1^n(U_1^n),W_2^n(U_2^n),X_1^n(W_1^n,\hat{h}_1^n,\hat{h}_2^n)$,
$X_2^n(W_2^n,\hat{h}_1^n,\hat{h}_2^n),Y^n) \in T_{\epsilon}^n\}
\rightarrow 1$ as $n \rightarrow \infty$. The error can occur
because of the following three events {\bf{E1}}-{\bf{E3}}. We show
that $P({\bf{Ei}}) \rightarrow 0$, for $i= 1,2,3$. For simplicity we
take $\delta=\epsilon$.

{\bf{E1}} The encoders do not find the codewords. However from rate
distortion theory~\cite{Cover04elements}, P. 356, \\ $\lim_{n \to
\infty}P(E_1)=0$ if $R_i^{'} > I (U_i;W_i), i \in {1,2}$.

{\bf{E2}} There exists another codeword $\hat{w}_1^n$ such that
$(\hat{w}_1^n,W_2^n,x_1^n(\hat{w}_1^n,\overline{h}_1^n,\overline{h}_2^n),x_2^n(W_2^n,\overline{h}_1^n,\overline{h}_2^n),Y^n)\in
T_\epsilon^n$. Define $ \alpha {\buildrel\Delta \over=}
(\hat{w}_1^n,W_2^n,x_1^n(\hat{w}_1^n,\overline{h}_1^n,\overline{h}_2^n),x_2^n(W_2^n,\overline{h}_1^n,\overline{h}_2^n),Y^n)$.
Then,
\begin{eqnarray}
P({\bf{E2}}) = P {\{\text{There is} ~ \hat{w}_1^n \ne w_1^n : \alpha
\in T_{\epsilon}^n}\}\\ \nonumber
 \leq \sum_{\hat{w}_1^n \ne W_1^n:(\hat{w}_1^n,W_2^n) \in T_{\epsilon} ^n} P{\{\alpha \in T_{\epsilon}^n\}}
 \label{e1}
\end{eqnarray}
Denote $\{(x_1^n(.),x_2^n(.),y^n):{\alpha \in T_{\epsilon}^n}\}$ by
$A$. The  probability term inside the summation in (\ref{e1}) is
\begin{gather*}
\leq\sum_AP\{x_1^n(\hat{w}_1^n,\overline{h}_1^n,\overline{h}_2^n),x_2^n(w_2^n,\overline{h}_1^n,\overline{h}_2^n),y^n|\hat{w}_1^n,w_2^n,\tilde{h}_1^n,\tilde{h}_2^n\}\\
\leq\sum_AP\{x_1^n(\hat{w}_1^n,\overline{h}_1^n,\overline{h}_2^n)|\hat{w}_1^n,\tilde{h}_1^n,\tilde{h}_2^n\}~~~~~~~~~~~~~~~~~~~~~~~~~~~~~~~~~~~~~~~~\\
P\{x_2^n(w_2^n,\overline{h}_1^n,\overline{h}_2^n),y^n|w_2^n,\tilde{h}_1^n,\tilde{h}_2^n\}\\
\leq\sum_A2^{-n\{H(X_1|W_1,\tilde{H}_1,\tilde{H}_2)+H(X_2,Y|W_2,\tilde{H}_1,\tilde{H}_2)-4\epsilon\}}~~~~~~~~~~~~~~~~~~~~~~~~~~~~~~~\\
\leq2^{n{H(X_1,X_2,Y|W_1,W_2,\tilde{H}_1,\tilde{H}_2)}}~~~~~~~~~~~~~~~~~~~~~~~~~~~~~~~~~~~~~~~~~~~~~~~\\
2^{-n\{H(X_1|W_1,\tilde{H}_1,\tilde{H}_2)+H(X_2,Y|W_2,\tilde{H}_1,\tilde{H}_2)-6\epsilon\}}\nonumber.
\label{e2}
\end{gather*}

But from hypothesis, we have
\begin{eqnarray*}
H(X_1,X_2,Y|W_1,W_2,\tilde{H}_1,\tilde{H}_2)-
H(X_1|W_1,\tilde{H}_1,\tilde{H}_2)\\\nonumber
-H(X_2,Y|W_2,\tilde{H}_1,\tilde{H}_2) \nonumber
= -I(X_1;Y|X_2,W_2,\tilde{H}_1,\tilde{H}_2).\nonumber \label{e3}
\end{eqnarray*}
Hence,
\begin{gather}
P\{(\hat{w}_1^n,W_2^n,x_1^n(\hat{w}_1^n,\overline{h}_1^n,\overline{h}_2^n),x_2^n(W_2^n,\overline{h}_1^n,\overline{h}_2^n),Y^n)\in
T_{\epsilon}^n\}\\\nonumber \leq
2^{-n\{I(X_1;Y|X_2,W_2,\tilde{H}_1,\tilde{H}_2)-6\epsilon\}}.
\label{e4}
\end{gather}
Then from \eqref{e1}
\begin{gather}
P({\bf{E2}})\leq \sum_{\hat{w}_1^n \ne w_1^n:(\hat{w}_1^n,w_2^n)\in T_{\epsilon}^n} 2^{-n\{I(X_1;Y|X_2,W_2,\tilde{H}_1,\tilde{H}_2)-6\epsilon\}}~~~~~~~~~~~~~~~~~~~\nonumber\\
 \leq |\{\hat{w}_1^n:(\hat{w}_1^n,w_2^n)\in T_{\epsilon}^n\}|
 2^{-n\{I(X_1;Y|X_2,W_2,\tilde{H}_1,\tilde{H}_2)-6\epsilon\}}\nonumber\\
 \leq |\{\hat{w}_1^n\}| P\{(\hat{w}_1^n,w_2^n)\in T_{\epsilon}^n\}
 2^{-n\{I(X_1;Y|X_2,W_2,\tilde{H}_1,\tilde{H}_2)-6\epsilon\}}\nonumber\\
 \leq 2^{n\{I(U_1;W_1)+\epsilon\}}2^{-n\{I(W_1;W_2)-\epsilon\}}~~~~~~~~~~~~~~~~~~~~~~~~~~~~\nonumber\\
 2^{-n\{I(X_1;Y|X_2,W_2,\tilde{H}_1,\tilde{H}_2)-6\epsilon\}}\nonumber\\
 =2^{n\{I(U_1;W_1|W_2)\}}2^{-n\{I(X_1;Y|X_2,W_2,\tilde{H}_1,\tilde{H}_2)-8\epsilon\}}.~~~~~~~~~~~~
\label{e5}
 \end{gather}
The R.H.S of the above inequality tends to zero if $ I(U_1;W_1|W_2)
< I(X_1;Y|X_2,W_2,\tilde{H}_1,\tilde{H}_2)$. In \eqref{e5} we have
used the fact that
\begin{eqnarray*}
I(U_1;W_1)-I(W_1;W_2)=I(U_1;W_1|W_2). \label{e6}
\end{eqnarray*}
Similarly, by symmetry of the problem we require
\begin{equation}
I(U_2;W_2|W_1) < I (X_2;Y|X_1,W_1,\tilde{H}_1,\tilde{H}_2).
\end{equation}
{\bf{E3}} There exist other codewords $\hat{w}_1^n$ and
$\hat{w}_2^n$  such that $\alpha {\buildrel\Delta\over
=}(\hat{w}_1^n,\hat{w}_2^n,x_1^n(\hat{w}_1^n,\overline{h}_1^n,\overline{h}_2^n),x_2^n(\hat{w}_2^n,\overline{h}_1^n,\overline{h}_2^n),y^n)\in
T_{\epsilon}^n$. 

Following the steps similar to that for  P({\bf{E2}}), P({\bf{E3}}) approaches zero for large $n$ if $
I(U_1,U_2;W_1,W_2) < I(X_1,X_2;Y|\tilde{H}_1,\tilde{H}_2)$.

Thus as $n \rightarrow \infty$, with probability tending to 1, the
decoder finds the correct sequence $(W_1^n,W_2^n)$ which is jointly
weakly $\epsilon$-typical with $(U_1^n,U_2^n)$.

The fact that $(W_1^n,W_2^n)$ is weakly $\epsilon$-typical with
$(U_1^n,U_2^n)$ does not guarantee that $f_D^n(W_1^n,W_2^n)$ will
satisfy the distortions $D_1,D_2$. For this, one needs that
$(W_1^n,W_2^n)$ are distortion-$\epsilon$-weakly typical
(\cite{Cover04elements}) with   $(U_1^n,U_2^n)$. Let
$T_{D,\epsilon}^n$  denote the set of -$\epsilon$-distortion typical sequences. Then by strong law of large numbers
$P(T_{D,\epsilon}^n|T_\epsilon^n)\rightarrow 1$ as $n\rightarrow
\infty$. Thus the distortion constraints are also satisfied by
$(W_1^n,W_2^n)$ obtained above with a probability tending to 1 as $n
\rightarrow \infty$. Therefore, if distortion measure $d_i$ is bounded
$\lim_{n \rightarrow \infty}E[d(U_i^n,\hat{U}_i^n)] \leq
D_i+\epsilon,~i=1,2$.

If there exist $u_i^*$ such that $E[d_i(U_i,u_i^*)]<\infty,~i = 1,2
$, then the result extends to unbounded distortion measures also as
follows. Whenever the decoded $(W_1^n,W_2^n)$ are not in the
distortion typical set then we estimate $(\hat{U}_1^n,\hat{U}_2^n)$
as $({u_1^*}^n,{u_2^*}^n)$. Then for $i=1,2$,
\begin{equation}
\label{ala} E[d_i(U_i^n,\hat{U}_i^n)] \leq D_i+\epsilon +
E[d(U_i^n,{u_i^*}^n) {\bf{1}}_{\{(T_{D,\epsilon}^n)^c\}}].
\end{equation}
Since $E[d(U_i^n,{u_i^*}^n)] < \infty $ and
$P[({T_{D,\epsilon}^n})^c] \rightarrow 0$ as $n \rightarrow \infty
$, the last term in RHS of  \eqref{ala} goes to zero as  $n \rightarrow
\infty$.

The above proof also hold for continuous alphabet sources and 
channels. The Markov lemma and weak typical decoding, the devices
used to prove the theorem continue to hold and the proof extends
with $E[X_i^2] \le \overline{P}_i~i=1,2$.
\end{appendix}

\scriptsize\parskip=10pt
\bibliographystyle{abbrv}
\bibliography{mybibfilefade}
\end{document}